\documentclass[conference]{IEEEtran}
\IEEEoverridecommandlockouts
\usepackage{cite}
\usepackage{amsmath,amssymb,amsfonts}
\usepackage{algorithmic}
\usepackage{graphicx}
\usepackage{textcomp}
\usepackage{xcolor}
\def\BibTeX{{\rm B\kern-.05em{\sc i\kern-.025em b}\kern-.08em
    T\kern-.1667em\lower.7ex\hbox{E}\kern-.125emX}}

\usepackage{tikz}
\usepackage{amsmath}
\usepackage{amssymb}
\usepackage{centernot} 
\usepackage{etoolbox} 
\usepackage{listings}
\usepackage{subcaption}
\usepackage{wrapfig}
\usepackage{enumitem}
\usepackage{listings}
\usepackage{hyperref}
\usepackage[capitalize,noabbrev,sort]{cleveref} 
\usepackage{url}
\usepackage{algorithm}
\usepackage{textcomp}
\usepackage{xspace}
\usepackage{wrapfig}
\usepackage{mdframed}
\usepackage{titlesec}
\usepackage{balance}
\usepackage[most]{tcolorbox}
\sloppypar

\newtoggle{comments}
\toggletrue{comments}

\newcommand{\reva}[1]{{{#1}}}
\newcommand{\revb}[1]{{{#1}}}
\newcommand{\revc}[1]{{{#1}}}

\newcommand{\reve}[1]{{{#1}}}
\newcommand{\revfail}[1]{{{#1}}}
\newcommand{\revname}[1]{{{#1}}}
\newcommand{\revexp}[1]{{{#1}}}
\newcommand{\revcons}[1]{{{#1}}}
\newcommand{\revfaileval}[1]{{{#1}}}

\newcommand{\camready}[1]{{{#1}}}
\newcommand{\appendixver}[2]{#1}

\newcommand{\ignore}[1]{}
\newcommand{\topic}[1]{}

\newcommand{\sysname}{ReCraft}
\newcommand{\sysnamelong}{\underline{rec}onfigurable \underline{Raft}}

\newcommand{\confold}{$C_{old}$}
\newcommand{\confoldnew}{$C_{old,new}$}
\newcommand{\confnew}{$C_{new}$}
\newcommand{\confnewl}{$C_{new-q}$}
\newcommand{\confsub}{$C_{sub}$}
\newcommand{\confsubs}{$C_{subs}$}
\newcommand{\confsubi}{$C_{sub.i}$}
\newcommand{\confjoint}{$C_{joint}$}
\newcommand{\conftx}{$C_{TX}$}
\newcommand{\confabort}{$C_{abort}$}

\newcommand{\qnew}{$Q_{new}$}
\newcommand{\qnewl}{$Q_{new-q}$}
\newcommand{\qold}{$Q_{old}$}

\newcommand{\splitenterjoint}{SplitEnterJoint}
\newcommand{\splitleavejoint}{SplitLeaveJoint}
\newcommand{\requestvote}{EnterElection}
\newcommand{\vote}{HandleVote}

\newcommand{\splitenterjointtt}{\texttt{\splitenterjoint{}}}
\newcommand{\splitleavejointtt}{\texttt{\splitleavejoint{}}}
\newcommand{\requestvotett}{\texttt{\requestvote{}}}
\newcommand{\votett}{\texttt{\vote{}}}

\newcommand{\mergeprepare}{MergePrepare}
\newcommand{\handlemergeprepare}{HandleMergePrepare}
\newcommand{\mergecommit}{MergeCommit}
\newcommand{\handlemergecommit}{HandleMergeCommit}

\newcommand{\handlemergepreparett}{\texttt{\handlemergeprepare{}}}

\newcommand{\preconenum}{\textbf{P\arabic*}}
\newcommand{\preconenumprime}{\textbf{P\arabic*'}}

\newcommand{\principleenum}{{\arabic*}.}

\newcommand{\boxparen}[1] 
{\begin{tcolorbox}[breakable, enhanced, boxsep=1pt, 
	left=1pt, right=1pt, top=2pt, bottom=2pt, 
	arc=0pt, boxrule=1pt, colback=white]{#1}
\end{tcolorbox}}
\newcommand{\enumparen}[1]{\begin{enumerate}[labelindent=2pt,  
noitemsep,topsep=0pt, left=5pt..18pt]{#1}\end{enumerate}}
\newcommand{\customenumparen}[3]{
	\begin{enumerate}[label={#1},noitemsep,topsep=0pt, left=5pt..25pt]
	\setcounter{enumi}{#2}
	{#3}
	\end{enumerate}
}

\newcommand{\itemparen}[1]{\begin{itemize}[labelindent=2pt, itemindent=0pt,
left=1em, noitemsep]{#1}\end{itemize}}

\newcommand{\para}[1]{\vspace{0.5ex}\noindent\textit{\textbf{#1}}}
\newcommand{\paranoskip}[1]{\noindent\textit{\textbf{#1}}}

\begin{document}

\title{\sysname{}: Self-Contained Split, Merge, and Membership Change of Raft Protocol}

\author{
\IEEEauthorblockN{
	Kezhi Xiong\textsuperscript{1} $\quad$
	Soonwon Moon\textsuperscript{2} $\quad$
	Joshua H. Kang\textsuperscript{1} $\quad$
	Bryant Curto\textsuperscript{1} $\quad$
	Jieung Kim\textsuperscript{3} $\quad$
	Ji-Yong Shin\textsuperscript{1}
}
\IEEEauthorblockA{
	\textit{
		\textsuperscript{1}Northeastern University$\quad$
		\textsuperscript{2}Seoul National University$\quad$
		\textsuperscript{3}Yonsei University$\quad$
	}
}
}
\maketitle
\thispagestyle{plain}
\pagestyle{plain}
\newtheorem{theorem}{Theorem}
\newtheorem{definition}{Definition}
\newtheorem{lemma}{Lemma}
\newtheorem{proof}{Proof}
\begin{abstract}
Designing reconfiguration schemes for consensus protocols is challenging
because subtle corner cases during reconfiguration could invalidate the 
correctness of the protocol. Thus, most systems that embed consensus protocols
conservatively implement the reconfiguration and 
refrain from developing an efficient scheme. Existing implementations 
often stop the entire system during reconfiguration and rely on a 
centralized coordinator, which can become a single point of failure. 
We present \sysname{}, a novel reconfiguration protocol for Raft, which supports 
multi- and single-cluster-level reconfigurations. 
\sysname{} does not rely on external coordinators and blocks minimally.
\sysname{} enables the sharding of Raft clusters with split and merge 
reconfigurations and adds a membership change scheme that 
improves
Raft. 
We prove the safety and liveness of \sysname{} and demonstrate its efficiency through implementations in etcd.

\end{abstract}


\section{Introduction}
\newcommand{\reconfigfootnote}{%
	\footnote{
		Configurations can include various information, 
		but our main interest is member nodes of a distributed system.
}%
} 
Reconfiguration\reconfigfootnote{} of a distributed system is vital to keep the
system scalable, cost-effective, and alive. For example, 
reorganizing the cluster depending on user loads can avoid service 
slowdowns or save operational costs, and timely replacement of failed nodes
can preserve the system's availability.  

The correctness of most distributed protocols relies on the configuration, and
it is crucial that reconfigurations are performed in a consistent 
and fault-tolerant manner. Problems with reconfigurations can invalidate 
the core protocol assumption
and stop the system~\cite{confbug, confbug2, cloud-bugs}. Thus, many 
systems 
manage the configuration using strongly consistent and highly 
available services, such as Chubby~\cite{chubby}, 
Zookeeper~\cite{hunt2010zookeeper}, and etcd~\cite{etcd}.

Then how do these strongly consistent systems manage and change their 
configurations?  Internally, these systems build on 
multi-Paxos~\cite{paxos-parliament, paxos-simple,paxos-moderate},
and Raft~\cite{raft} which are  
consensus-based state machine replication (SMR)~\cite{SMR} protocols that are
logically equivalent~\cite{vivaladifference}. 
Reconfiguration schemes for consensus protocols can be categorized by whether
they rely on an external configuration manager and whether the system must block
during the reconfiguration process. 
Chubby~\cite{chubby} and vertical Paxos~\cite{paxos-vertical} assume an 
external configuration manager, which can be another multi-Paxos cluster. 
Most other Paxos variants manage the configuration within 
itself but require the reconfiguration command to be \emph{committed} through 
consensus before execution~\cite{paxos-moderate, stoppablepaxos, reconfiguringpaxos, 
generalizedconsensus, wormspace, smart, vrrevist}. Raft improves them 
in a \emph{wait-free} fashion which optimistically 
applies the new configuration right after the command 
is \emph{received}~\cite{raft, ongarothesis}. 


While reconfiguration schemes for multi-Paxos and Raft systems are capable of 
changing member nodes of a single cluster instance, they cannot fully address 
the multi-cluster reconfiguration that relates to scalability.
Multi-Paxos and Raft can scale for the read throughput by adding passive learner
nodes, but the write throughput does not scale by adding nodes.
Consensus-based SMR protocols are designed to maintain identical copies of data 
typically through a single leader multicasting to all other members. Thus, 
adding more nodes only decreases the write throughput. Existing solutions to scale
the write throughput include employing multiple leader nodes~\cite{mencius},
logically partitioning physical nodes to concurrently commit different sets of
commands~\cite{scaling-paxos}, and decoupling log ordering and
replication~\cite{epaxos}. Still, these approaches focus on maintaining a single
instance of the SMR cluster and do not support scaling out to many. Raft-based
storage systems like TiKV~\cite{tidb} and
CockroachDB~\cite{cockroachdb} support scaling beyond a single cluster by
managing multiple logical sharded instances of Raft. However, they require a 
coordinator module outside of Raft, which can be a single point of 
failure. Also, the protocol is designed conservatively to avoid potential bugs:
cluster reconfigurations run for a maximum of two clusters at a time in sequential steps with 
frequent blocking.  

In this paper, we present \sysname{}, a \sysnamelong{}, which augments Raft
with 1) a \emph{self-contained} consensus-based concurrent cluster split and merge
protocol that does not rely on any external service and 2) a new membership
change protocol that is more fault-tolerant.

With \sysname{}, a Raft instance can split into two or more instances that
handle disjoint data sets, and multiple instances can merge into one. The split
and merge decisions are made by the consensus of all participating nodes. The
protocol mixes push and pull-based communication to ensure that disconnected 
nodes or clusters during the reconfiguration can catch up. 
The protocol looks simple, like the original Paxos or Raft protocol, which can 
be summarized in a few lines of pseudo code. However, \sysname{} is a result of 
iterative revisions and formal proofs to handle subtle corner cases. 

The \sysname{} membership change alters multiple member nodes and quorum sizes
in a single step in a wait-free fashion. The scheme relies on a recent formal
proof of generalized Raft reconfiguration~\cite{adore}. \sysname{} instantiates
the general theory into a practical multi-node reconfiguration design and
implementation. Compared to the Raft joint consensus reconfiguration, it is
simpler to reason and exhibits better fault tolerance for practical cluster
sizes as it requires fewer votes.


\topic{Challenging to get membership change right}
To ensure that \sysname{} protocols are 
bug-free and compatible with the main Raft protocol, we present proof of 
safety and liveness under the same assumption as Raft
(Section~\ref{sec:proof}). 
For safety properties, we prove that \sysname{} preserves the immutability and 
linearizability of its state (i.e., state machine safety) under reconfigurations. 
For the liveness, we prove that the split and merge protocols keep all nodes and 
clusters alive. 
Due to the logical equivalence of Raft and 
multi-Paxos~\cite{raftstar, ado}, our protocol can also be ported to
multi-Paxos.


We implement \sysname{} in etcd~\cite{etcd} and its Raft library, which many other
systems rely on, including TiKV and CockroachDB.
Evaluations of \sysname{} on a public cloud and a theoretical 
analysis against etcd/TiKV/CockroachDB demonstrate that \sysname{} is safe, live, and efficient. 

This paper makes the following contributions:
\itemparen {
	\item To our knowledge, the first self-contained split/merge protocol 
		for Raft and Paxos protocols.
	\item A new wait-free membership change protocol that improves   
        Raft for practical cluster sizes.
	\item Formal safety and liveness proofs of \sysname{} protocols. 
	\item The implementation of \sysname{} in etcd
		and an evaluation in a cloud environment. 
}

\section{Background}
\label{sec:background}


In this section, we introduce Raft~\cite{raft} and its
reconfiguration~\cite{ongarothesis} which are the basis 
of \sysname{} and discuss split/merge protocols for Raft in TiKV and CockroachDB.

\subsection{Raft}

\subsubsection{Basic Workings} 
Raft replicates the SMR log using distributed consensus in a strongly consistent 
manner to a cluster of nodes in two stages: leader election and log replication.

\para{Leader Election.} Raft uses the monotonically increasing term 
and its SMR log records log entries with the term when the entry is generated. 
There is at most one elected leader per term. The leader constantly sends log 
replication requests or heartbeat messages to
non-leader nodes or followers. Each follower maintains a random timer that
resets if it hears from the leader. If the timer goes off, the follower turns
into a candidate, increments the term number, and multicasts a leader election
message for the term to all nodes in the cluster. The election message contains
the term number and the recency of the candidate's local SMR log. Each node has
one vote per term and sends its vote to the candidate who first contacts with
the same or later log state than itself. The candidate that gathers the votes 
from a majority quorum of the cluster becomes the leader. If none of the 
candidates receives the majority votes, the leader election reoccurs in the 
next term. 

\para{Log Replication.} All replication requests for log entries go through the 
leader. Similar to the leader election, the leader multicasts the log entry 
with the current term number to all nodes in the cluster. The followers
append the entry only if the leader's term number is equal to or greater 
than the follower-perceived term and their logs match with the leader's log
up to where the new log entry is about to be appended. In case a follower's log
is out of sync, the leader first keeps the follower's log up-to-date and then 
appends the new log entry. If a majority quorum accepts the replication request, 
the log entry is committed and becomes immutable. 

\para{Quorum Overlap.} 
The majority quorum support in each step ensures that the next leader is 
supported by at least one node that committed the last committed log entry and 
the elected leader's log contains all committed log entries. This quorum overlap 
is one of the key properties that lead to the safety guarantees
of the protocol. Both Raft and \sysname{} reconfiguration protocols heavily 
rely on the quorum overlap.

\subsubsection{Membership Changes}

\begin{figure}
\includegraphics{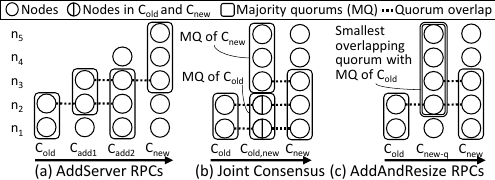}
\caption{Reconfiguring a 2-node cluster (\confold{}) to a 5-node cluster
(\confnew{}) using Raft and \sysname{} reconfiguration schemes.}
\label{fig:reconfig}
	\vspace{-1em}
\end{figure}

Raft employs two types of membership change schemes: the Add/RemoveServer RPC
(AR-RPC) for a single member node change in one consensus
step~\cite{ongarothesis} and the joint consensus (JC) for arbitrary node changes
in two consensus steps~\cite{raft}. Both are wait-free in which nodes
optimistically apply the configuration change immediately after receiving it 
as a special SMR log entry and converge to the committed configuration.

Both approaches leverage the quorum overlap property between old and new 
configurations. The AR-RPC (Figure~\ref{fig:reconfig}a) adds or removes
one node at a time, and the one-node difference naturally maintains the quorum 
overlap: all majority quorums (MQ) of an old $N$ node configuration 
\confold{} always overlap with those of new  $N+1$ or $N-1$ node 
configuration \confnew{}. On the other hand, the JC (Figure~\ref{fig:reconfig}b) 
induces the quorum overlap and works in two steps. It starts with the old 
configuration \confold{} and enters the joint mode by committing the joint 
configuration \confoldnew{} to quorums of \confold{}. Under the joint mode, any 
decision should be made by the quorums of both old and new configurations, and 
these quorums subsume those of \confold{}. Then, the new configuration
\confnew{} is committed to a quorum of the new configuration, and the 
reconfiguration completes. Similarly, the quorums of \confoldnew{} subsume 
those of \confnew{}, so the quorum overlap is always maintained.

\subsection{Generalized Add/RemoveServer RPC}
\label{subsec:bugandgen}

The single-step AR-RPC is simpler and faster than the JC but can only alter one 
node at a time. The restriction comes from preconditions to fulfill before 
starting the AR-RPC:
\customenumparen{\preconenum{}}{0}{
	\item All prior reconfiguration requests in the leader's log must be
		committed.
	\item The new configuration should differ from the current 
		configuration by at most one node.
	\item The leader must commit a log entry in its term before starting a new 
		reconfiguration.
}

Among them, \textbf{P3} was later added to deactivate concurrent reconfigurations
that led to a subtle bug~\cite{raft-bug}. While formally verifying that
\textbf{P3} indeed removes the bug, Honor\'{e} et al.~\cite{adore} showed that
\textbf{P2} can be relaxed to, 
\customenumparen{\preconenumprime{}}{1}{
	\item Consecutive configurations should always maintain the quorum overlap.
}
They formally verified the safety of any reconfiguration scheme that satisfies 
the preconditions with this generalization, including the AR-RPC.  
The \sysname{} membership change protocol is an instance that satisfies the
safety proof and sits between the AR-RPC and the JC 
(Section~\ref{sec:membershipchange}).

\subsection{Cluster Split/Merge in Multi-Raft Designs}
\label{subsec:tikvsm}

TiKV and CockroachDB implement multi-Raft, which is multiple sharded 
clusters of Raft. Both systems employ similar splitting and merging 
mechanisms using an external cluster manager (CM) that drives the 
operation as follows~\cite{cockroachdbsm}.

\para{Split.} To split cluster $C_{src}$, the CM commits a 
new subrange command to $C_{src}$. The CM brings up cluster $C_{dst}$ that 
will take the remaining subrange, and data is copied from $C_{src}$ to 
$C_{dst}$. After the copying completes, $C_{dst}$ starts the service. 

\para{Merge.} To merge clusters $C_{src}$ and $C_{dst}$, the CM stops 
$C_{src}$ by committing a special command and copies data from $C_{src}$ to 
$C_{dst}$. Once the data is fully copied, the CM increases $C_{dst}$'s range to 
include $C_{src}$'s range.

Split and merge operations of TiKV/CockroachDB can be viewed as the CM moving a 
subset of data and range from one cluster to a new cluster and migrating the 
entire data and range of a cluster to another cluster, respectively.
While the centralized CM handling the split and merge simplifies the high-level 
process, interleaved or concurrent reconfigurations can complicate the
operations. Thus, the CM uses distributed locks/transactions 
at the layers above to solve these problems. 

%



\section{Splitting and Merging \sysname{} Clusters}

Most production systems running consensus-based SMR implement key-value
interfaces~\cite{hunt2010zookeeper, etcd, mongodbprod}, and independent access
to keys naturally lends the system to sharding as in distributed hash
tables~\cite{chord, can, pastry, tapestry}. Based on our goal to design a
consistent, failure-resilient, and live split and merge protocol for Raft 
clusters, we establish two design principles:
\customenumparen{\principleenum{}}{0}{ 
\item Rely minimally on external services or 
manual controls. 
\item Utilize existing consensus mechanisms in Raft that guarantee strong
consistency and high availability. 
}

\subsection{Preliminaries}

\paranoskip{Preconditions}
Splitting and merging reconfigurations of \sysname{} are decided by the 
consensus of the member nodes similar to Raft. \sysname{} applies the same 
preconditions \textbf{P1}, \textbf{P2'}, and \textbf{P3} for all 
reconfigurations to prevent outdated reconfiguration commands potentially coming 
from multiple clusters from interfering with the up-to-date one. 

\revfail{\paranoskip{Failure Assumptions}
\sysname{} shares the same failure model as Raft and Paxos: it assumes an
asynchronous network and non-Byzantine node failures.}
Each \sysname{} cluster, which we interchangeably refer to as  
a configuration $C$, can tolerate $f = n-q$ node failures, where 
$n$ is the number of nodes in the cluster and $q$ is the quorum size. 
$q$ is typically the majority in Raft and Paxos, but during intermediate steps
of \sysname{} reconfigurations, $q$ can temporarily grow larger than the 
majority but never smaller \revfail{like Raft.
\sysname{} always guarantees safety 
(Theorem~\ref{thm:state-machine-safety}) under the failure model. However, 
for liveness, \sysname{}, like Raft, assumes at least a quorum of nodes---which 
may vary depending on the reconfiguration phase---in each cluster reacts to requests 
within a finite number of retrials. Both for safety and liveness,
any node (i.e., no exceptions for leader nodes) can fail at any time as long as 
these assumptions hold.} \sysname{} does not 
rely on external services except for a naming service in a rare case 
(Section~\ref{sec:recovery}). Thus, the protocol is self-contained and there is 
no single point of failure even during reconfiguration, unlike other SMR 
systems with sharding like TiKV/CockroachDB~\cite{tidb, cloud-bugs}. 

\begin{figure*}
\scriptsize
\lstdefinelanguage{mypseudo}{ 
	escapeinside={(*}{*)},
	morecomment=[l]{//},
	morecomment=[s]{/*}{*/},
    string=[b]",
	keywords={if, else, foreach, in, return},
    basicstyle={\linespread{0.7}\ttfamily}, 
	commentstyle=\itshape,
	keywordstyle={\color{blue}},
	keywordstyle=[2]{\color{red}},
	morekeywords=[2]{appendEntry, notifyCommit, requestVote, pullLog,
	respondPull}
}
\lstset{
numbers=left, numberstyle=\tiny, stepnumber=1, numbersep=5pt,
showstringspaces=false, mathescape=true
}
\hspace{0.02\textwidth}
\begin{minipage}[t]{0.32\textwidth}
\begin{lstlisting}[language=mypseudo, firstnumber=1]
// Leader - enter joint consensus
bool (*\textbf{\splitenterjointtt{}}*)(Config (*\confjoint{}*)) {
  votes = [];
  $C$ = getCurrentConfig();
  // Precondition check
  if (!P1 || !P2' || !P3) 
    return FAILURE;
  foreach (n in memberNodes($C$)) { 
    if (appendEntry(n, (*\confjoint{}*)) == OK) 
      votes.add(n); 
  }
  applyElectConfig((*\confjoint{}*)); 
  if (isQuorum($C$, votes)) return 
    SUCCESS;
  else return FAILURE;
} 
// Leader - leave joint consensus
bool (*\textbf{\splitleavejointtt{}}*)(Config (*\confnew{}*)) {
  votes = [];
\end{lstlisting}
\end{minipage}
\begin{minipage}[t]{0.31\textwidth}
\begin{lstlisting}[language=mypseudo, firstnumber=20]
  $C$ = getCurrentConfig();
  if (!isJoint($C$)||!isCommitted($C$)) 
    return FAILURE;
  foreach (n in memberNodes($C$)) {
    if (appendEntry(n, (*\confnew{}*)) == OK) 
      votes.add(n); 
  }
  (*\confsub{}*) = myConfig((*\confnew{}*));
  applyCommitConfig((*\confsub{}*));
  if (isQuorum((*\confsub{}*), votes)) {
    notifyCommit((*\confold{}*), (*\confnew{}*));
    applyElectConfig((*\confsub{}*));
    IncEpoch();
    return SUCCESS;
  } else return FAILURE;
}
// Candidate
bool (*\textbf{\requestvotett{}}*)() {
  $C$ = getCurrentConfig();
\end{lstlisting}
\end{minipage}
\begin{minipage}[t]{0.33\textwidth}
\begin{lstlisting}[language=mypseudo, firstnumber=39]
  foreach (n in memberNodes($C$)) { 
    res = requestVote(n)
    if (res == OK) votes.add(n);
    if (res == PULL) { 
      pullLog(n); 
      return FAILURE; 
  } }
  if (isQuorum($C$, votes)) 
    return SUCCESS;
  else return FAILURE;
}
// Follower - response to request vote
void (*\textbf{\votett{}}*)(Request req) {
  $E_{curr}$ = getCurrentEpoch();
  $E_{req}$ = getEpoch(req.term);
  if ($E_{curr}$ > $E_{req}$) 
    respondPull(req.node); 
  else ...  /*same as Raft voting*/ 
}
\end{lstlisting}
\end{minipage}
\vspace{-0.1in}
  \caption{%
	  Pseudo code of functions for the split protocol. \reve{Functions in red
	include communications: appendEntry, requestVote and PullLog are 1-to-1
	RPCs, notifyCommit is a multicast RPC, and respondPull sends
	an acknoweldgment.}
  }%
  \label{fig:splitpseudo}
\end{figure*}


\begin{figure*}
	\vspace{-1em}
    \includegraphics{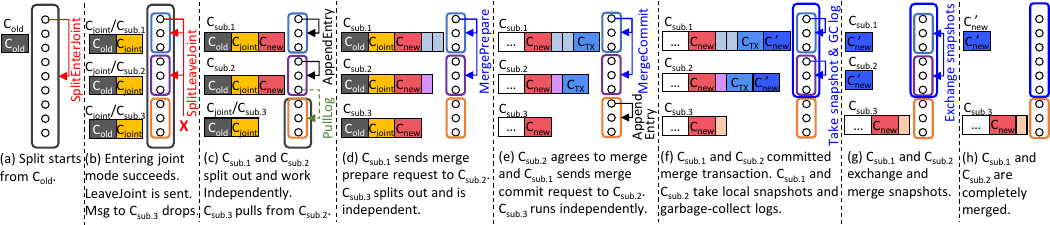}
        \caption{An example of a series of split and merge operations from a 
cluster-level viewpoint. Rounded rectangles show clusters' node configurations
and cluster's committed log states are on the left. \confold{} splits to
    $C_{sub.1-3}$ (steps a-d), and $C_{sub.1}$ and $C_{sub.2}$ merge to
    $C'_{new}$ (steps d-h).}
        \label{fig:splitmergeeg}
	\vspace{-1em}
\end{figure*}

\para{Epoch Numbers.} 
We introduce monotonically increasing epoch numbers to order the configurations 
committed for splitting and merging. We add the epoch number as a prefix to 
Raft's term number (e.g., the first 4 bytes as the epoch number and the
remainder as the regular term number for an 8-byte integer). The epoch numbers are 
located at higher digits of the new term number, so an updated epoch number for 
a split/merge will prevent commands from old configurations from interfering 
with the new configuration. Epoch numbers also work as indicators for nodes that 
failed to participate in the reconfiguration to realize that their peers have 
moved on. Note that epoch numbers are not updated for membership change 
reconfigurations within a single cluster.

\newcommand{\subclusterfootnote}{%
\footnote{We use the term subcluster to refer to clusters
generated from a cluster split, or clusters merging to form a single 
cluster.}%
}

\subsection{Split}
\label{sec:split}
\sysname{} can split a cluster into multiple 
subclusters\subclusterfootnote{}. They share the same SMR log of the original
cluster but evolve independently after the split. The overlay layer, such as 
the Zookeeper or etcd, can selectively accept requests for the subcluster's 
new range or redirect them to appropriate subclusters.

We propose a variant of the JC for split operations so that all subclusters 
have clues of the ongoing split even if they go offline in the middle. 
During the split, we use different election and log commit quorums.
The pseudo code of our two-phased split protocol is in 
Figure~\ref{fig:splitpseudo} and an example is in Figure~\ref{fig:splitmergeeg}.
Note that the return of ``SUCCESS'' in the 
code means the operation fully 
succeeded, but ``FAILURE'' means the operation may or may not have  
failed and requires a re-execution (e.g., a leader committing 
log entries from past terms in Raft).  


\para{Configurations.} 
The new configuration \confnew{} should include how many subclusters the  
current configuration \confold{} will split into and each new subcluster $i$'s 
configuration \confsubi{}. Each \confsubi{} includes disjoint members and 
data ranges of \confold{}.
\confjoint{} is the intermediate step to reach \confnew{} and has the same
information as \confnew{}. \confjoint{} only changes how the election quorum 
works from \confold{}: majority quorum votes from each and every \confsubi{} is 
required. 
When \confnew{} arrives at each node, the node extracts its own \confsubi{} from 
\confnew{} and applies it. 

\para{\splitenterjoint{}.} 
To initiate a split, the leader proposes entering the joint mode with
\confjoint{} (Figure~\ref{fig:splitmergeeg}a). 
The preconditions \textbf{P1}, \textbf{P2'}, and \textbf{P3} 
must be satisfied (line 6) and \confjoint{} is sent to the member nodes as a 
special log entry that applies immediately after \textit{receiving.} 
The leader applies the configuration in a wait-free fashion regardless of the 
quorum acknowledgment (line 12). Applying the configuration only changes the election
quorum and the quorum for log commit still uses \confold{}. Until 
\confjoint{} is committed, the leader cannot propose to leave the joint mode, 
but it can propose regular log entries. 

\para{\splitleavejoint{}.}
To leave \confjoint{}, the leader first confirms that \confjoint{} is
\textit{committed} (line 21). Then it sends \confnew{} to all members of
\confjoint{} (Figure~\ref{fig:splitmergeeg}b) and applies \confsub{} in a
wait-free fashion (lines 23-28). However, \confsub{} is not fully applied at
this stage: the leader communicates with nodes in \confsub{} for committing
\confnew{} and log entries that come after; communicates with all nodes in
\confold{} to replicate \confnew{} and entries that come before; and uses the 
quorum of \confjoint{} to elect a new leader, if necessary. After confirming the 
commit of \confnew{}, the leader notifies the commit of \confnew{} --- 
similar to how a committed index of regular log entry gets notified in Raft --- 
to all members of the \confold{} (line 29-30), fully applies \confsub{} (line 31), 
and updates the epoch number (line 32).  
From this point on, the split for \confsub{} of the leader is completed and 
\confsub{} can run independently of others. 

The follower applies \confsub{} the same way: once it receives 
\confnew{} it only changes the commit quorum, and the election quorum updates 
after confirming the commit of \confnew{}. The commit notification (line 30) is 
necessary so that candidates from other subclusters, if they have \confnew{} in 
their log, can know of its commit and elect a leader within its subcluster. 
If subclusters do not have the \confnew{} log entry, they
have to first catch up with the log (described below).

\para{Pulling through \revb{\requestvote{}} and \vote{}.} In the event of a network 
disruption during the last phase of splitting, some nodes or even an entire 
subcluster could miss the message from \splitleavejoint{}. If the split 
succeeds for a \confsub{} (i.e., \confnew{} is committed to the quorum of 
\confsub{}), nodes in \confsub{} that missed the message will be eventually 
contacted by their peers and catch up with the configuration. 
However, if an entire subcluster misses the message to leave the joint mode and 
other subclusters have moved forward, it has no way to make progress under a 
regular Raft-like setting. 
The members of the subcluster are in \confjoint{} and they need quorum votes from 
all other subclusters to elect a leader. However, other subclusters will have 
more recent logs 
and higher epoch numbers, so they will not respond to the missed-out subcluster.

\begin{figure*}
\scriptsize
\lstdefinelanguage{mypseudo}{ 
	escapeinside={(*}{*)},
	morecomment=[l]{//},
	morecomment=[s]{/*}{*/},
    string=[b]",
	keywords={if, else, foreach, in, return},
    basicstyle={\linespread{0.7}\ttfamily}, 
	commentstyle=\itshape,
	keywordstyle={\color{blue}},
	keywordstyle=[2]{\color{red}},
	morekeywords=[2]{appendEntryToConf, Prepare2PC, respond, Commit2PC},
}
\lstset{
numbers=left, numberstyle=\tiny, stepnumber=1, numbersep=5pt,
showstringspaces=false, mathescape=true
}
\hspace{0.02\textwidth}
\begin{minipage}[t]{0.33\textwidth}
\begin{lstlisting}[language=mypseudo, firstnumber=1]
// Coordinator cluster leader 
bool (*\textbf{\mergeprepare{}}*)(Config (*\conftx{}*)) {
  2pcVotes = []; votes = [];
  $C$ = getCurrentConfig();
  // Precondition check
  if (!P1 || !P2' || !P3) 
    return FAILURE;
  // Local TX decision to own cluster
  (*\conftx{}'*) = setDecision((*\conftx{}*), OK);
  votes = appendEntryToConf($C$, (*\conftx{}'*));
  // send 2PC prepare to (*\confsub*)
  foreach ((*\confsub*) in clusters((*\conftx{}*))) { 
    if (Prepare2PC((*\confsub*), (*\conftx{}*)) == OK)
      2pcVotes.add((*\confsub*));   
  }
  if (isQuorum($C$, votes) && 
      isUnanimous((*\conftx*), 2pcVotes)) {
    applyConfig((*\conftx{}'*)); 
    return SUCCESS;
  } else return FAILURE;
}
// Non-coordinator cluster leader
bool (*\textbf{\handlemergepreparett{}}*)(Config (*\conftx{}*)) {
\end{lstlisting}
\end{minipage}
\begin{minipage}[t]{0.33\textwidth}
\begin{lstlisting}[language=mypseudo, firstnumber=24]
  votes = [];
  $C$ = getCurrentConfig();
  // Precondition check
  if (!P1 || !P2' || !P3) res = NO;
  else res = OK;
  // Local TX decision to own cluster
  (*\conftx{}'*) = setDecision((*\conftx{}*), res);
  votes = appendEntryToConf($C$, (*\conftx{}'*));
  if (isQuorum($C$, votes)) {
    applyConfig((*\conftx{}'*));
    respond(res);
    return SUCCESS;
  } else return FAILURE;
}
// Coordinator cluster leader
bool (*\textbf{\mergecommit{}}*)(bool res, (*\confnew{}*)) {
  2pcAcks = []; 
  votes = [];
  $C$ = getCurrentConfig();
  if (!underTX($C$)||!TXPrepared((*\conftx{}*)))
    return FAILURE;
  // Final TX decision to own cluster
  $C_{next}$ = (res == COMMIT)?(*\confnew*):(*\confabort*);
\end{lstlisting}
\end{minipage}
\begin{minipage}[t]{0.32\textwidth}
\begin{lstlisting}[language=mypseudo, firstnumber=47]
  votes = appendEntryToConf($C$, $C_{next}$);
  // send 2PC commit/abort to (*\confsub*)
  foreach ((*\confsub*) in clusters((*\conftx{}*))) { 
    if (Commit2PC((*\confsub*), $C_{next}$) == OK)
      2pcAcks.add((*\confsub*));   
  }
  if (isQuorum($C$, votes) &&
      allResponded((*\conftx*), 2pcAcks)) { 
    applyConfig($C_{next}$);
    return SUCCESS;
  } else return FAILURE;
}
// Non-coordinator cluster leader
bool (*\textbf{\handlemergecommit{}}*)(Config $C_{next}$) {
  votes = [];
  $C$ = getCurrentConfig();
  votes = appendEntryToConf($C$, $C_{next}$);
  if (isQuorum($C$, votes)) {
    applyConfig($C_{next}$);
    respond(OK);
    return SUCCESS;
  } else return FAILURE; 
}
\end{lstlisting}
\end{minipage}
\vspace{-0.1in}
  \caption{%
	Pseudo code of functions for the two-phase commit of the merge protocol. 
	\reve{Functions in red
	include communications: Prepare2PC, and Commit2PC are 1-to-1 RPCs, 
	appendEntryToConf is a multicast RPC, and respond sends an
	acknowledgement.}
  }%
  \label{fig:mergepseudo}
  \vspace{-1em}
\end{figure*}
%


\sysname{} adds a pull mechanism so that the missed-out subcluster 
($C_{sub.3}$ in Figure~\ref{fig:splitmergeeg}b) can save itself. 
The missed-out subcluster must first learn about the split 
completion and it should catch up with the log status up to the entry of 
\confnew{}. \revb{All \sysname{} nodes can respond to the node that sends vote requests 
with a lower epoch number to pull log entries} from themselves (line 55). 
The node receiving the pull response pulls the \textit{committed} log entries 
from the responder (line 43 and Figure~\ref{fig:splitmergeeg}c). \reva{This is 
similar to the routine log-catchup operations during the appendEntry RPC of Raft 
in reverse directions which does not incur high-overhead.} Once 
the \confnew{} entry is pulled, the subcluster applies \confsub{} and elects 
its leader.

Epoch numbers play two critical roles here: 1) they prevent missed-out nodes of 
other subclusters from turning leaders of up-to-date subclusters into followers 
with a large term number, and 2) they clearly flag the commit of \confnew{} so 
that the missed-out nodes can confidently pull log entries. 
Uncommitted log entries can be overwritten so a node in \confnew{} halfway 
(i.e., applied \confnew{} as the commit configuration but not election 
configuration) 
\revb{should not respond to pull requests. However, without the epoch number, 
whether the node has fully moved on to \confnew{} or not is unclear from 
external observers' point of view. The epoch number clearly marks the commit of 
\confnew{} and from which nodes should the missed-out nodes pull.}


\para{Differences from the Raft Joint Consensus.} Unlike Raft's joint 
configuration \confoldnew{}, \sysname{}'s \confjoint{} does not explicitly 
require quorums from \confold{} and \confnew{} for election and commit. 
\confjoint{} only requires quorums of all \confsubi{} for election as they are 
guaranteed to overlap with the quorum of \confold{}. 

Using different quorums for leader election and log commit is also very 
different from Raft. As quorums of \confold{} and \confjoint{} overlap, it is
safe to commit to \confold{} under the joint mode. The quorum of \confold{} is
equal to or smaller than that of \confjoint{} so this expedites commit. 
Using the quorums of \confjoint{} for election until completely committing 
\confnew{} ensures all leaders elected until the split completes share the 
same view of the log up to \confnew{} entry; immediately reducing the election 
quorum size can violate quorum overlap properties for log entries between 
\confjoint{} and \confnew{} and break safety guarantees.
The use of different quorum sizes for election and commit resembles Flexible 
Paxos~\cite{flexiblepaxos}, but \sysname{} uses this idea primarily to
guarantee safety among concurrent subclusters. 


Pulling log entries from other subclusters is a fully new feature 
in \sysname{}. It resembles a Paxos-style leader election where proposers 
collect up-to-date data from other nodes or learner nodes that collect
committed states from others.

\para{Subtle Corner Cases.} During the pull operations, the data source
node may not have all the necessary log entries or may even have a more outdated 
log than the puller. If the source node was also a missed-out node and only 
received a request vote message from an up-to-date node with the updated epoch 
number, then its term/epoch number will become 
up-to-date but its log state could still be outdated. Since the pull operation 
only copies committed log entries, it does not break safety. The puller 
can contact different nodes for the latest data or wait for the outdated node 
to be updated.

Epoch numbers are updated only after confirming the commit of \confnew{} instead
of when partially applying \confnew{} after receiving it. We initially 
thought of fully applying \confnew{} and increasing the epoch number with the 
configuration. However, 1) as explained above, the safety can break for
immediately changing the election quorum size, and 2) it made the epoch 
number grow arbitrarily large for a single split if \confnew{} fails to 
commit multiple times; this confuses the nodes to needlessly pull data
from each other. The current design ensures safety and clearly marks a split 
success with the epoch increase.


\subsection{Merge}
\label{sec:merge}
 
The \sysname{} merge protocol consolidates multiple clusters that manage
disjoint data into one. The merge is more involved than the split,
as it synchronizes and unifies states of multiple subclusters. 
The merge protocol consists of three phases (Figure~\ref{fig:splitmergeeg}d-h): 
the first two phases match
each step of the two-phase commit (2PC) protocol to make the merge decision 
as a transaction, and the third phase exchanges
subcluster states to create a common ground to resume as a single  
cluster. The data exchange phase blocks because the data 
transfer takes time, and at least a quorum of nodes in the merged cluster  
must be in the same state to preserve safety properties for leader election
and log commits. Unlike the split 
that extends the JC, the merge adds new mechanisms to Raft.

\para{Configurations and Overview.} 
The new configuration of the merger \confnew{} holds the members of
the resulting cluster and from which subcluster \confsubi{} they come. 
\confnew{} also includes the combined data range, a unique transaction id, and 
which subcluster is coordinating the merger. To reach \confnew{}, the 
subclusters must go through the 2PC, where the transaction decision depends on 
all merging subclusters \confsubi{}'s. As the first phase of 
the 2PC, all subclusters receive \conftx{}, which includes 
\confnew{} as the transaction intent. Individual subclusters decide 
whether to join the merger or not and commit \conftx{} with the local 
decision. As the second phase of the 2PC, if all clusters 
agree to merge, then \confnew{} is committed to all clusters; otherwise, 
\confabort{}, which nullifies \conftx{}, is committed. Between committing 
\conftx{} and \confnew{}/\confabort{}, each cluster behaves the same, servicing 
regular client requests of the original configuration. Once \confnew{} is 
committed, the data exchange happens among the subclusters. Finally, the merged 
cluster elects a new leader and resumes as a single cluster. 

\subsubsection{Distributed Transactions for Merge Decisions}
The 2PC protocol for the merge runs at the cluster level (pseudo code in 
Figure~\ref{fig:mergepseudo}), and it begins when the 
merge request arrives at one of the merging subclusters. The entire 
subcluster that was contacted becomes the coordinator to drive the 2PC. 
All \sysname{} nodes have the logic to drive the 2PC and the leader of the
subcluster takes the lead: the coordinator is naturally as robust as the Raft
cluster unlike TiKV/CockroachDB's cluster manager.  

\para{\mergeprepare{}.} 
The leader of the coordinator subcluster starts the merger by triggering 
\mergeprepare{} (Figure~\ref{fig:splitmergeeg}d).
Similar to the split, it checks the three preconditions for reconfiguration
(lines 6) and commits the local ``OK'' decision to its own cluster 
(lines 8-10). Then, it sends the 2PC prepare request to all involved clusters
(lines 11-15). If the local commit succeeds and all clusters acknowledge 
``OK'' then the \mergeprepare{} succeeds (lines 16-19) and can move on to 
committing \confnew{}. If any of the subclusters acknowledged ``NO,'' then the 
merger fails, and the next step is to commit \confabort{}. 

\para{\handlemergeprepare{}.}
The non-coordinator subclusters treat the 2PC prepare request similarly to a 
regular client request and the leader of the subcluster handles the request 
with \handlemergeprepare{}. 
The main reason for the ``NO'' vote is \textbf{P1}, an ongoing reconfiguration, 
as \textbf{P2'} is guaranteed by the protocol and \textbf{P3} can be easily 
fulfilled by committing a no-op log entry (lines 27-28). Even when the cluster 
votes ``NO,'' the decision must be recorded for proper execution of the 2PC 
protocol (lines 30-31). The response to the coordinator subcluster is sent  
after the decision is committed (lines 32-36). 

\para{\mergecommit{} and \handlemergecommit.} 
Once the transaction decision is finalized, the commit/abort phase of the 
2PC takes place (Figure~\ref{fig:splitmergeeg}e). 
If the decision is an ``ABORT'' then the coordinator sends \confabort{} to all
subclusters; otherwise, \confnew{} is sent (lines 45-52). Reconfigurations 
\confabort{}/\confnew{} must be applied after it is 
\textit{committed} to the local subcluster (lines 60-65). 
This is required, because nodes successfully applying \confnew{} immediately move
on to the data exchange phase. The coordinator cluster applies the configuration 
last after checking that all subclusters completed the 2PC (lines 53-56).

\revb{\para{Handling Failures.} 
The merge issues subcluster-granularity operations and as long as each 
subcluster is alive (i.e., satisfies our failure assumption for a cluster), 
the merge operation goes through. For example, a complicated failure 
scenario is the leader node of the coordinating cluster failing. In this case, 
recovery schemes for Raft and 2PC are used (not in Figure~\ref{fig:mergepseudo}). 
With the Raft recovery, the log entries related to the ongoing transaction will 
be noticed and committed to the local subcluster, if necessary, by the new 
leader. Then, the new leader will learn from the committed logs which stage of 
the 2PC execution was interrupted. 2PC transactions are designed to be
idempotent using unique ids, and the new leader can resume the 
2PC from the last known successful state.}

\subsubsection{Data Exchange and Resumption.}
After the merge decision is finalized, the members of the merged cluster should 
resume from the same SMR log state. 

\para{Data Exchange.}
There can be many ways to merge subcluster states, such as concatenating or
merge-sorting the logs. However, we choose to take snapshots of the replayed 
log \revb{(i.e., the key-value map of the etcd layer)}, exchange them, and use 
the combined snapshot as the base state for the merged cluster. We adopt the 
snapshot exchange because the snapshot is generally smaller than the log
\revb{(e.g., the log can contain multiple updates to the same key but the
snapshot of the replayed log only remembers the last update).}

Once \confnew{} is committed, the snapshot exchange takes place. 
Applying \confnew{} includes creating a snapshot of the local log up 
to the log entry before \confnew{}, trimming the log, 
and pulling the snapshots from other subclusters to create the combined 
snapshot (i.e., a key-value store state with combined key ranges). 
With the snapshot, nodes in the merged cluster start fresh with the 
log that begins with the \confnew{} entry (Figure~\ref{fig:splitmergeeg}f-g). 
Note that log entries in subclusters that come after the 
\confnew{} entry are discarded; logs are committed in a sequence, so 
it is safe to discard these uncommitted entries.

\revcons{
The basic \sysname{} consistency guarantee is linearizability 
(or strict serializability of key-value pairs in the etcd context) on 
each cluster. However, the merge combines multiple linearized data chunks that 
are partially ordered. Thus, we regard the merged cluster---which inherits the 
ordering of each subcluster---as the start of a new cluster where linearizable 
updates begin; this is the same as TiKV and CockroachDB. 
While the current implementation only deals with disjoint data chunks, if the 
application (e.g., the etcd layer) running on \sysname{} requires special 
ordering guarantees across merging data, additional logic can be implemented 
in the application.}

\para{Resumption.} 
Before resuming operations, the merged cluster must decide the new epoch and 
term numbers to use. Before the commit of \confnew{} and the exchange of 
snapshots, each subcluster uses its own epoch and term numbers. During the 
first phase of the 2PC, the coordinator collects the epoch numbers of all 
subclusters, and includes a new epoch number $E_{new}$ greater than the maximum 
$E_{max}$ of all subclusters in \confnew{}: i.e., $E_{new} = E_{max} + 1$. 
After \confnew{} is applied, \confnew{} entry is treated as committed at term 
0 of epoch $E_{new}$. 
 
The first leader election of the merged cluster starts at term 1 of epoch 
$E_{new}$. However, not all nodes of \confnew{} may be in sync when the first 
leader election occurs. Still, it is guaranteed that all subclusters have 
committed \confnew{}: a candidate with $E_{new}$ can arise only after merging 
the snapshots from all subclusters, and exchangeable snapshots become 
available after committing \confnew{}. In a sense, the merge is a reverse 
process of the split where the 2PC mimics the joint consensus mode in which 
every subcluster has to approve the merge decision. For the merged cluster to 
carry on in \confnew{} with data from all subclusters, it is sufficient to use 
the majority quorum. Note that nodes that are unaware of \confnew{} cannot 
become the leader due to the small epoch number and outdated log.  

\para{Handling Missed-out Nodes.} The merge protocol only requires a
majority quorum from each subcluster to complete the merger, and there can be
nodes that are behind. Unlike the split, the up-to-date log of the merged 
cluster looks very different from that of old subclusters, since the log is 
garbage collected and the cluster state relies on the merged snapshot. 

\sysname{} employs two ways to keep the nodes up-to-date.
First, the missed-out nodes can update themselves by contacting other nodes that
have higher epoch numbers using \requestvote{} and pullLog functions that are used  
in the split process. Second, the leader node that knows the 
entire reconfiguration history and monitors the follower's log status during 
appendEntry RPC call can install the snapshot and missing log entries. 
This is similar to Raft's InstallSnapshot RPC~\cite{ongarothesis}
but works for nodes coming from different subclusters. 

\para{Resizing the Merged Cluster.}
The merged cluster naturally holds a large number of member nodes, and it is
likely that the cluster size needs to be reduced. The membership change 
reconfiguration after the merge works, but the resizing can be done while 
applying \confnew{}. The \sysname{} merge protocol resembles the stoppable 
Paxos~\cite{stoppablepaxos}, where a cluster temporarily stops for the 
reconfiguration. 
Thus, it is possible to select a subset of nodes in
\confnew{} to resume as the merged cluster. However, an arbitrary node selection
can lead to choosing only missed-out nodes that are outdated. 
The safety requirement is to guarantee overlap between the combined quorum of all \confsubs{}
and the quorum of the resized cluster: e.g., selecting all members of one or more 
\confsubs{} as the resized cluster fulfills this. Such member information can be added to \confnew{}.


\section{\sysname{} Membership Change}
\label{sec:membershipchange}

The membership change of a single cluster is another important reconfiguration
for the scalability and longevity of a cluster. 
We propose a variant of the AR-RPC and the JC for the single cluster 
membership change based on the generalized proof surrounding 
\textbf{P2'}~\cite{adore} (Section~\ref{subsec:bugandgen}). 
Similar to vanilla Raft reconfigurations, the proposed scheme is wait-free and
always maintains quorum overlaps between consecutive configurations. \sysname{} 
can add/remove multiple nodes to/from a cluster at once, but unlike the 
JC that requires votes from a specific set of nodes (i.e., from old and new 
configurations), \sysname{} does not distinguish the source. 

\subsection{Add/RemoveAndResize and ResizeQuorum}
\para{Adding Nodes.} \sysname{} offers AddAndResize RPC to add $n$ new nodes to 
configuration \confold{} that has $N_{old}$ members. The RPC adds $n$ nodes to 
the cluster, modifies the quorum size from \qold{} to 
\qnewl{} $= N_{old} + n - Q_{old} + 1$, and places the cluster into an 
intermediate configuration \confnewl{} in a single consensus step 
(Figure~\ref{fig:reconfig}c).
``New-q'' indicates that the configuration has the same members as the 
final configuration \confnew{} but uses a different quorum size \qnewl{} for both leader
election and log commit.  \qnewl{} is the smallest quorum size that 
forces quorums of \confold{} and \confnewl{} to overlap.

The RPC adds $n$ nodes to \confold{} in a single step, but the quorum size 
\qnewl{} needs to be adjusted to the majority for the cluster to work like a 
regular Raft cluster. To reach this final configuration \confnew{}, the 
ResizeQuorum RPC is used to reset the quorum size to \qnew{} $= \lfloor
\frac{N_{new}}{2} \rfloor+1$ in a single consensus step. \confnewl{} and 
\confnew{} share the same members, and \qnewl{} is always equal to or greater 
than \qnew{}, so the quorums of \confnewl{} and \confnew{} are guaranteed to 
overlap. Thus, the safety of ResizeQuorum RPC holds. 
 

\para{Removing nodes.} Removing nodes from the cluster works similarly using 
the RemoveAndResize RPC to reach \confnewl{} and then calling ResizeQuorum RPC 
to reach the target \confnew{} with the majority quorum. The quorum size in 
\confnewl{} during the removal is adjusted to \revb{\qnewl{} $= N_{old} - Q_{new}
+ 1$}
to force the quorum overlap between \confold{} and \confnewl{}.

Unlike the AddAndResize RPC, which can add an unbounded number of nodes, the 
RemoveAndResize RPC can remove up to $r < Q_{old}$ nodes from \confold{}. 
Because the maximum quorum size of \confnewl{} is capped by the cluster size 
$N_{new-q}$, if $r$ becomes equal to or greater than \qold{}, the quorum 
overlap between \confold{} and \confnewl{} cannot be satisfied.

\subsection{Comparisons with Raft's Membership Changes}
\label{subsec:membercomparison}

\para{Number of Consensus Steps.}
When adding or removing one node, the AR-RPC performs better than the 
JC, as the AR-RPC requires one consensus step. \sysname{} works the same as 
the AR-RPC as one node difference makes \qnewl{} and \qnew{} to be equal
and obviates the ResizeQuorum RPC.
When adding or removing two nodes, \sysname{} performs the best.
The AR-RPC must be called twice to add/remove two nodes, which requires the 
same two consensus steps as the JC. \sysname{} can handle adding two nodes in 
a single step when \confold{} has an even number of nodes and \qnewl{} equals 
\qnew{}. Otherwise, \sysname{} takes the same two consensus steps as the JC. 
Beyond altering two nodes, \sysname{} takes the same number of consensus steps 
as the JC. However, if the cluster size needs to be reduced by more than half, 
which is rare in practice, then \sysname{} requires extra steps. 

\begin{figure}
\centering
\includegraphics{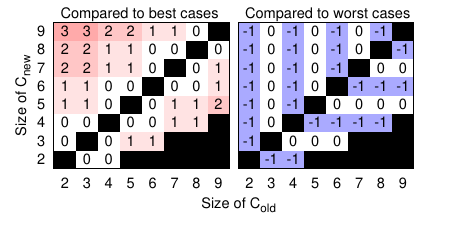}
    \vspace{-0.5em}
\caption{The number of additional votes \sysname{} requires during intermediate 
steps of the membership change compared to the best cases and the worst cases
of the JC.}
\label{fig:memberchangeeval}
    \vspace{-1em}
\end{figure}

\para{Number of Necessary Vote Messages and Fault Tolerance.} 
Beyond the number of consensus steps, how do \sysname{} and the JC
differ? The main difference is the number of votes to collect to commit a log 
entry in the intermediate configurations. \sysname{} needs votes from a quorum 
larger than the majority (i.e., \qnewl{}) and the JC requires majority quorum 
votes from both old and new configurations. For example, 
in Figure~\ref{fig:reconfig}, \sysname{} always needs votes from at least four 
nodes in \confnewl{}, but the JC under \confoldnew{} requires two votes from 
\confold{}$=\{n_1, n_2\}$ and three votes from \confnew{}$=\{n_1, n_2, n_3, n_4,
n_5\}$. Depending on which order the votes arrive at the leader, the JC 
can commit a log entry with three to five votes. $n_1$ and $n_2$ belong
to both \confold{} and \confnew{}, so their votes count twice for each 
configuration. If their votes arrive first, one additional vote can fulfill 
the joint quorum. However, if votes of $n_3$, $n_4$, and $n_5$ arrive first, 
and the leader should wait for votes of $n_1$ and $n_2$ to fulfill the quorum 
of \confold{}. For adding and removing nodes, the number of necessary votes for 
the best case is $V_{best} = max(Q_{new}, Q_{old})$, and the worst case is 
$V_{worst} = |N_{new} - N_{old}| + min(Q_{new}, Q_{old})$. 


Necessary vote counts relate to fault tolerance and tail latency  
if there are stragglers. In Figure~\ref{fig:reconfig}, \sysname{} can tolerate 
any one node failure under \confnewl{}. The JC under \confoldnew{} can tolerate 
two node failures out of $\{n_3, n_4, n_5\}$, which is better than 
\sysname{}, but it cannot tolerate any failure of $\{n_1, n_2\}$, which makes 
\sysname{} better. In essence, \sysname{} conservatively sets the quorum size 
for the quorum overlap to treat all nodes equally. 
\revb{Still, 
\sysname{} requires the same or fewer votes than $V_{worst}$ of the JC. 
}

The cells in Figure~\ref{fig:memberchangeeval} show how many more votes 
\sysname{} needs than the JC when reconfiguring a cluster of size 
on the x-axis to a new size on the y-axis. 
Positive numbers in red mean that the JC requires fewer votes, and negative 
numbers in blue mean \sysname{} requires fewer votes. 
\revb{Compared to the best cases for the JC (left) when votes for the
nodes in both old and new configurations arrive first, the JC 
always requires fewer votes than \sysname{} when adding or removing more than 
two nodes. However, for more practical cases of adding or removing one or two
nodes, \sysname{} and JC perform comparably. \sysname{} requires the same number 
of votes for altering one node and the same or one more votes for altering two 
nodes.
Compared to the worst cases for the JC (right), \sysname{} always requires 
the same or fewer votes.}


Overall, the required votes for \sysname{} and the JC are comparable, 
but for configuring member nodes between the most widely used 
\revb{Raft cluster sizes of 3 and 5, \sysname{} is a more appealing choice for 
the number of consensus steps as the fault tolerance is comparable.}
Additionally, \sysname{}'s membership change protocol 
is simpler to reason and program as it treats all nodes equally.



\section{Recovery from Long-Term Failures}
\label{sec:recovery}

Continuous split, merge, and membership changes can confuse  
nodes that were partitioned for an extended period. Following our first design 
principle of relying minimally on external parties, \sysname{} employs means to 
restore these nodes.

\para{Restoring a Node.} 
Restoring the offline node when it gets back online is simple. 
When any of the live, up-to-date nodes perceive this node as their cluster 
member, they will eventually contact and update the node. Alternatively, if the 
peer nodes in the disconnected node are still online, the node can contact 
them to pull up-to-date log entries from them and catch up. 

\para{Restoring a Cluster.}
Restoring a long-term offline cluster is more subtle. There could be cases
where the cluster missed \splitleavejoint{} call during a cluster split. If the
peer clusters are around, the pull-based recovery approach in 
Section~\ref{sec:split} suffices. However, 1) all peer clusters could have 
merged with other clusters, garbage collected the logs, and no longer have 
records that the offline cluster needs, or more radically
2) all members of the peer clusters could have been removed. 

To handle the first case, \sysname{} requires all clusters to maintain the 
reconfiguration history even after garbage collecting the log. Thus, even if 
clusters have gone through a series of reconfigurations, the offline cluster 
can restore itself from nodes that were in the same configuration. 
The reconfiguration history is garbage collected after confirming that all 
nodes involved have completed the reconfiguration.

For the second case, 
\sysname{} uses a naming service that maintains the information of all 
live clusters. All clusters register their configuration to the naming service
after reconfigurations. The naming service needs to be consistent with the 
cluster with a very loose time bound like the domain name service (DNS). 
Through the service, the cluster can find nodes that share 
the reconfiguration history, pull the history, and restore itself. This is the only time 
\sysname{} relies on an external service. 

\revname{Note that in a practical deployment, distributed systems commonly 
leverage a DNS-like naming service so that clients can find 
individual systems and bootstrap (e.g., Chubby, Google's etcd-like service 
relies on the naming service~\cite{chubby}). Similarly, \sysname{} relies on 
this service for liveness, but we can instead make successor nodes of old 
configurations to monitor and restore all left-out nodes. Although this would 
implement a truly self-contained protocol, it adds unnecessary complexity to 
handle a very rare failure scenario.}

\newcommand{\disj}{\mathbin{\#}}
\newcommand{\ndisj}{\mathbin{\centernot\#}}
\newcommand{\lastTerm}{\operatorname{lastTerm}}
\newcommand{\length}{\operatorname{length}}

\section{\sysname{} Safety and Liveness Proofs}
\label{sec:proof}

To guarantee the safety and liveness of \sysname{}, we offer relevant 
proofs for split and merge. We only describe the main components of the 
proof and the remaining proof is available \appendixver{in the Appendix}{as a technical report~\cite{anonurl}}. 
We omit the proof for membership changes, as the proof by 
Honor\'{e} et al.~\cite{adore} subsumes it.

\subsection{Safety} 
The safety proof for \sysname{} follows a similar structure to existing
Raft protocol proofs~\cite{raftext, cppraft}, and the top-level 
theorem is stated as: 

\boxparen{
	\begin{theorem}[State machine safety]\label{thm:state-machine-safety}
		If a node has applied a log entry at a given index to its state 
		machine, no other node will ever apply a different log entry for 
		the same index.
	\end{theorem}
}

To prove Theorem~\ref{thm:state-machine-safety}, we start by 
defining analogous properties of \sysname{} to those of Raft
(Definitions~\ref{def:leader-append-only}-\ref{def:leader-completeness})
and adjusting sub-lemmas to introduce epoch numbers and multiple clusters.

\boxparen {
	\begin{definition}[Leader Append-Only]\label{def:leader-append-only}
	  A leader never overwrites or deletes its log entries and only appends.
	\end{definition}

	\begin{definition}[Election Safety]\label{def:election-safety}
	  At most one leader can be elected 
	  \emph{in a given epoch and term per cluster}.
	\end{definition}

	\begin{definition}[Log Matching]\label{def:log-matching}
		If two nodes of \emph{the same epoch and cluster configuration}
	  contain an entry with the same index and term in their logs,
	  then the logs of that epoch are identical in all entries up through 
	  the given index.
	\end{definition}

	\begin{definition}[Leader Completeness]\label{def:leader-completeness}
	  Given an epoch number $e$ and a cluster $C$,
	  if an entry is directly committed in a term of $(C,e)$,
	  then that entry will be present in the logs of the leaders
	  for all higher-numbered terms of $(C,e)$.
	\end{definition}
}

We also introduce a new definition of consensus and quorums
to distinguish different kinds of
consensus in \sysname{}:

\boxparen{
	\begin{definition}[Consensuses and quorums]\label{def:consensus-and-quorums}
	  There are three kinds of consensuses used in \sysname{}.
	\begin{itemize}[labelindent=1pt, itemindent=0pt, left=0em, noitemsep]
	    \item \textit{Normal consensus} for a cluster $C$ requires its 
		    majority.
	    \item \textit{Joint consensus} for a set of subclusters $C_1, \ldots, C_n$
		    requires majorities for each $C_1, \ldots, C_n$.
	    \item \textit{Constituent consensus} for a set of clusters $C_1, \ldots,
		    C_n$ requires a majority for one of $C_1, \ldots, C_n$.
	\end{itemize}
	  A quorum for a consensus $c$ is a specific set of nodes constituting the consensus.
	\end{definition}
}

Demonstrating that \sysname{} satisfies analogous lemmas to those of 
Raft is largely straightforward by following 
Raft's proof, though it necessitates modifications due to changes in
\sysname{}. For instance, split and merge operations introduce new
constraints beyond those relied upon by the original Raft: 
\enumparen {
\item A \confnew{} entry cannot be appended to the log until the \confjoint{} 
	entry is committed.
\item The candidate holding the \confjoint{} entry in its log should use 
	\confjoint{} for its election.
\item The leader under \confjoint{}/\conftx{} cannot propose a new configuration 
	other than the corresponding \confnew{}/\confabort{}.
}

These modifications require defining properties related to the cluster
reconfiguration of \sysname{}, as in
Definition~\ref{def:cluster-well-formedness}. It asserts that any
cluster split or merge will yield properly formed clusters
ensuring the safety of our cluster configuration.

\boxparen{
	\begin{definition}[Cluster Well-Formedness]\label{def:cluster-well-formedness}
	  Given an epoch number $e$ and given two clusters $C_1$ and $C_2$ of epoch $e$,
	  $C_1$ and $C_2$ are either identical or disjoint.
	\end{definition}
}

Another key characteristic of \sysname{} is relevant to log updates as 
in Definition~\ref{def:log-consistency}. It ensures that all
committed entries are unique in terms of their index, epoch, and cluster.

\boxparen{
	\begin{definition}[Log Consistency]\label{def:log-consistency}
	  Given an epoch number $e$ and a cluster $C$,
	  if two entries are committed at the same index in any member of $(C,e)$,
	  then they are indeed equal.
	\end{definition}
}

Reflecting the modifications in \sysname{}, we design our proofs in three
levels. Instead of proving all properties directly through induction on
epoch and term numbers, we focused on, 
\enumparen {
	\item Directly proving the induction on the step execution of \sysname{}
		for properties independent of the increase in term and
		epoch numbers (e.g, Definition~\ref{def:leader-append-only}).
	\item Demonstrating sub-lemmas (e.g., Definitions~\ref{def:election-safety},
		\ref{def:log-matching}, and \ref{def:leader-completeness})
		within a single epoch number based on the induction of term
		numbers. The key strategy for proving these properties is to
		establish that Definition \ref{def:cluster-well-formedness} always holds
		and then prove the
		aforementioned sub-lemmas as corollaries.
	\item The top-level safety theorem 
		(Theorem~\ref{thm:state-machine-safety}) can be derived from the
		corollary lemma (Definition~\ref{def:log-consistency}), which
		can also be proven by combining the stated sub-lemmas.
}

We focus on describing proof strategies for 
Definition~\ref{def:leader-completeness}, one of the most non-obvious
theorems, rather than outlining all property proofs. It 
significantly alters original Raft's definition due to the introduction of 
cluster and epoch. To simplify the proof, we leverage 
Definition~\ref{def:cluster-well-formedness} to show 
that all the cooperating nodes of that epoch share the same cluster configuration.

For example, consider a situation where a follower $a$ of epoch $e$ and cluster
configuration $C_a$ accepted a message from a leader $p$ with cluster
configuration $C_p$. Then $a$ and $p$ share the epoch number since otherwise $a$
wouldn't have accepted the message.  Moreover, $C_p$ contains $a$ as its member
since a leader can't send a message to a non-member.
Then $C_a$ and $C_p$ have at least one common element, $a$.
Then by the Cluster Well-Formedness, we can conclude that $C_a = C_p$.
With this in mind, we prove Lemma~\ref{lem:leader-completeness}.

\boxparen{
\begin{lemma}[Leader Completeness]\label{lem:leader-completeness}
  Given an epoch number $e$, the Cluster Well-Formedness of epoch $e$ 
  (Definition~\ref{def:cluster-well-formedness}) implies the Leader Completeness 
  of epoch $e$ (Definition~\ref{def:leader-completeness}.)
\end{lemma}
\begin{proof}
  By assuming the Definition~\ref{def:cluster-well-formedness} of epoch $e$,
  we ignore any concerns about cluster configuration while proving 
  Lemma~\ref{lem:leader-completeness}. In the following proof, we assume all 
  nodes are members of the same cluster $(C,e)$.

  Suppose
    a log entry $i_0 \mapsto (x_0, t_0)$ is directly committed by a leader $p_0$ at
    term $t_0$ and a leader $p_1$ is elected at $t_1$ where $t_0 < t_1$.
  Let $log_0$ be the log of $p_0$ at the time of committing $i_0 \mapsto (x_0, t_0)$,
  and $log_1$ be the log of $p_1$ at the time of its election.
  We have to show that $i_0 \mapsto (x_0, t_0)$ is contained in $log_1$.
  We use lexicographic strong induction on 
	$(i_0, t_1)$.

  Let $(c_0, q_0)$ be the consensus and the quorum used for committing 
  $i_0 \mapsto (x_0, t_0)$ and 
  $(c_1, q_1)$ be the consensus and the quorum used for the leader election of 
  $p_1 @ t_1$.
  $c_0$ can be one of \textit{normal} or \textit{constituent consensus}, while 
  $c_1$ can be one of \textit{normal} or \textit{joint consensus}.
  We claim that $q_0$ and $q_1$ are not disjoint. Use case analysis on $c_0$.
\begin{description}[leftmargin=1em]
    \item[Case 1:] If $c_0$ is a \textit{normal consensus}, we can immediately 
	show that $q_0$ and $q_1$ are not disjoint.
    \item[Case 2:] If $c_0$ is a \textit{constituent consensus}, 
	it means that there is a \confnew{} entry in $log_0$.
	In this case, there must exist a leader who created the \confnew{} entry.
	With this knowledge, there must be a committed \confjoint{} entry at the 
	time of creating the \confnew{} entry. Consider a directly committed 
	entry $i' \mapsto (x', t')$ which leads to the commit of the 
	\confjoint{} entry. In this case, $i' < i_0$.     
	Therefore, by the induction hypothesis with $i' \mapsto (x', t')$ and 
	$p_1 @ t_1$, it can be derived that $log_1$ contains $i' \mapsto (x', t')$.           
	By Definition~\ref{def:log-matching}, $log_1$ also contains the 
	\confjoint{} message. Hence, $p_1 @ t_1$ must use joint consensus for i
	ts election: i.e., $c_1$ is a \textit{joint consensus}. Thus, we 
	conclude that $q_0$ and $q_1$ are not disjoint.
  \end{description}

  Let $v$ be a node of $q_0 \cap q_1$ and $log_v$ be the log of $v$ 
  at the time of voting to $p_1$. We claim that $log_v$ contains 
  $i_0 \mapsto (x_0, t_0)$. By using the induction hypothesis with 
  $i_0 \mapsto (x_0, t_0)$ and leaders for the term between $t_0$ and $t_1$, 
  we know that all the intervening leaders contained $i_0 \mapsto (x_0, t_0)$.
  $v$ accepted $i_0 \mapsto (x_0, t_0)$ at term $t_0$ so it also had the entry.
  Since followers only remove entries if they conflict with the leader, we can 
  conclude that $i_0 \mapsto (x_0, t_0)$ would have preserved until $v$'s vote 
  to $p_1$.
  Since $v$ granted its vote to $p_1 @ t_1$, there are two cases to consider.
	\begin{description}[leftmargin=1em]
    \item[Case 1:]
      When $\lastTerm(log_v) < \lastTerm(log_1)$,
      assume that $i' \mapsto (x', t')$ is the last entry of $log_1$.
      Since $t_0 \le \lastTerm(log_v) < \lastTerm(log_1) = t'$, we can use the 
      induction hypothesis with $i_0 \mapsto (x_0, t_0)$ and the leader of $t'$.
      Then the leader of $t'$ must have contained $i_0 \mapsto (x_0, t_0)$ at 
      the time of election. By Definition~\ref{def:leader-append-only}, the log 
      of the leader of $t'$ at the time of creating $i' \mapsto (x', t')$ also 
      contains $i_0 \mapsto (x_0, t_0)$. Then, according to 
      Definition~\ref{def:log-matching}, we can conclude that $log_1$ contains 
      the entry $i_0 \mapsto (x_0, t_0)$.
    \item[Case 2:]
      When $\lastTerm(log_v) = \lastTerm(log_1)$ and 
      $\length(log_v) < \length(log_1)$, $log_v$ is a prefix of $log_1$. 
      Hence, $log_v$ also contains $i_0 \mapsto (x_0, t_0)$.
  \end{description}
Therefore, Definition~\ref{def:leader-completeness} (Leader Completeness) holds under the assumption of Definition~\ref{def:cluster-well-formedness} (Cluster Well-Formedness).  
\end{proof}
}

\subsection{Liveness} 
Any deterministic consensus protocols including \sysname{}, Raft and Paxos 
cannot guarantee liveness under unrestricted network failures~\cite{flp}, so
%
%
we show the liveness of \sysname{} based on the same assumption as
Raft. We assume message delivery to a quorum within a finite amount of retrials
and 
\begin{center}
$broadcastTime << electionTimeout << MTBF$.
\end{center}
In this context, $broadcastTime$ refers to the average duration for nodes
calling RPCs and receiving responses, while $electionTimeout$ represents
the duration between heartbeat messages. $MTBF$ stands for the mean time between
failures, which is the average duration between failures for a single node. 
With these assumptions in place, the liveness of \sysname{} is stated
in Theorem~\ref{thm:liveness}. 
\boxparen{
\begin{theorem}[Liveness]\label{thm:liveness}
\sysname{} responds to all requests from clients in a timely manner. 
\end{theorem}

\begin{proof}
Proving Theorem~\ref{thm:liveness} requires four case analyses. 
\enumparen{
\item Termination of entry appends by leaders of each cluster.
\item Termination of the normal leader election process,
\item Progress assurances for subclusters following a split,
\item Termination of a merge when subclusters combine.
}

\begin{description}[leftmargin=1em]

\item[Case 1, 2:] They are trivial, following the original Raft process and the
aforementioned assumptions. 

\item[Case 3:] This requires additional analysis during the split operation: when 
the leave split operation is acknowledged and committed in all subclusters, and
when it is missed in some subclusters. In the former case, all subclusters
needing leader election can immediately start the election. In
the latter case, nodes in the missed subclusters will send leader
election messages to known peers after the election timeout, which is also a part of the original Raft protocol.
This will trigger a pull-based recovery. If all known peers are unavailable, the nodes query
the naming service, which we assume always available, to find a node to pull 
from. 

\item[Case 4:] Due to the 2PC-style process of our
merging operation, as long as the majority of nodes in each subclusters are 
reachable and alive as in our assumption, proving liveness becomes 
straightforward.  
\end{description}
\end{proof}
}
\subsection{Proof Mechanization}\label{subsec:proof-mech}
To mitigate the risk of human error, we have utilized Rocq prover~\cite{rocq} to 
outline and validate our high-level proof structure. This approach provides a stronger
guarantee of the proof's correctness. \camready{
We wrote the mechanized proof structure of \sysname{} in Rocq from scratch, but
similar to our proof, Raft-related parts follow the structure of existing Rocq 
proofs that verify Raft~\cite{adore, cppraft}.
}


\section{Evaluation}


We implement \sysname{} in etcd version 3.5 by modifying 4K+ lines of Go code
in the etcd-Raft~\cite{etcd-raft} library. 
We evaluate \sysname{} on a public research cloud \revb{in a single datacenter
region} using up to 16 virtual
machine (VM) instances, each with 2 vCPUs and 8GB memory, for running etcd
and clients. We focus on the write
performance, which is our main interest, since the read performance scales
easily by adding learners~\cite{paxos-moderate}. The block drives of VMs
run on Cinder on Ceph, which performs poorly for small writes, so we use 
512B requests. 

Because etcd does not support split/merge, we compare \sysname{} against 
an emulated split and merge of TiKV/CockroachDB on etcd, which we call TC. 
We use emulation because both TiKV and CockroachDB only support splitting 
logical instances of Raft, do not provide control over the physical 
locations of logical Raft instances, cannot configure Raft instances 
to have different numbers of nodes, and do not allow three or more clusters
split/merge. We realize TC using a script that issues a series of commands 
from etcd admin tools~\cite{disaster-recovery} to go through the same steps 
as TiKV/CockraochDB split and merge and achieves the same outcome as \sysname{}.
Note that the split and merge of logical Raft clusters in TiKV/CockroachDB can
happen within the same physical nodes. \revexp{In this case, TiKV/CockroachDB can
transfer data more efficiently,} but we assume a case where Raft
clusters need to be in disjoint physical nodes for reliability and 
performance. 

\revc{All experiments are measured at least three times after 30 seconds of warm
up time from either the client (\cref{fig:etcdoverhead},
\cref{fig:splittime}, and \cref{fig:mergetime}) or the server driving
the reconfiguration (\cref{fig:splitlatency} and \cref{fig:mergelatency}).}

\subsection{Performance Overhead}
\begin{wrapfigure}[11]{r}{1.4in}
	\vspace{-2em}
\includegraphics{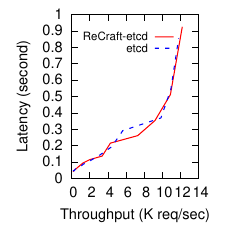}
\caption{etcd performance with \sysname{} vs Raft.}
\label{fig:etcdoverhead}
\end{wrapfigure}
\sysname{} adds new features to etcd, and we evaluate if \sysname{} adds any
overhead to regular etcd operations. Figure~\ref{fig:etcdoverhead} compares
the write performance of 3-node \sysname{}-etcd and unmodified etcd 
clusters as we increase the load. Both systems perform 
identically, showing that \sysname{} implementation does not affect the 
regular workings of etcd. 


\subsection{Split Performance}
\label{subsec:evalsplit}
\begin{figure}
	\centering
	\begin{subfigure}[b]{0.49\columnwidth}
		\centering
		\includegraphics{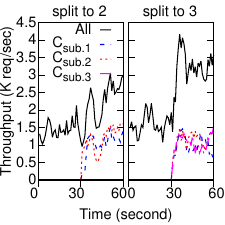}
		\vspace{-0.6em}
		\caption{Throughput}
		\label{fig:splittime}
	\end{subfigure}
	\begin{subfigure}[b]{0.49\columnwidth}
		\centering
		\includegraphics{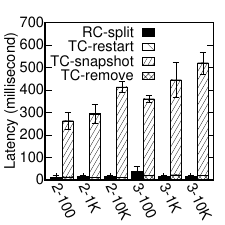}
		\vspace{-0.6em}
		\caption{Latency}
		\label{fig:splitlatency}
	\end{subfigure}
	\caption{Split performance. (a) Throughput of 6-node and 9-node cluster
	splitting into two (left) and three (right) subclusters, respectively.
	(b) Split latency for \sysname{} (RC) and TiKV/CockroachDB emulation
	(TC).  The x-tic, $a$-$b$, indicates a cluster with $b$ KV pairs
	splitting $a$-ways.} 
	\vspace{-1em}
\end{figure}

\para{Throughput.} Split operations require two consensus steps and do not add
noticeable overhead. Figure~\ref{fig:splittime} shows the throughput of 6-node
(left) and 9-node (right) \sysname{}-etcd instances splitting into two and three
3-node subclusters, respectively. We use 128 clients stressing the
cluster with a uniform random puts. There is no performance
impact when the split occurs at the 30-second mark. After the
split, subclusters divide the load and double and triple the combined
throughput.

\reva{We do not compare the throughput of \sysname{} against TC, as the 
throughput before and after the split/merge for the two systems is identical 
as they share the same etcd base. The only difference in the throughput 
measurement is the duration of the throughput dip when the split/merge occurs,
which is captured by the latency measurement below.}

\para{Latency.} We compare the latency for the split operation against TC. 
TC removes nodes that need to split through a membership change 
(e.g., 3 nodes are removed to split a 6-node cluster into two subclusters), 
takes a snapshot of the existing data inside removed nodes, installs snapshot 
and the subcluster configuration to the nodes, and restarts them as subclusters. 

We use the same node configurations as the throughput measurement but with etcd
instances holding 100, 1K, and 10K key-value (KV) pairs. \sysname{} takes almost
constant time to complete the split as it always entails committing two log
entries (Figure~\ref{fig:splitlatency}). However, TC must take multiple
steps as described above. TC takes on average 21x longer mainly due to the data 
migration. Excluding the data migration, the TC runs similarly to \sysname{} but 
TC's split is driven by a single external service which is also a potential single point of failure. 

\begin{figure}
	\centering
	\begin{subfigure}[b]{0.49\columnwidth}
		\centering
		\includegraphics{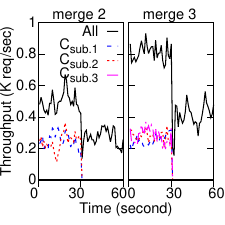}
		\vspace{-0.6em}
		\caption{Throughput}
		\label{fig:mergetime}
	\end{subfigure}
	\begin{subfigure}[b]{0.49\columnwidth}
		\centering
		\includegraphics{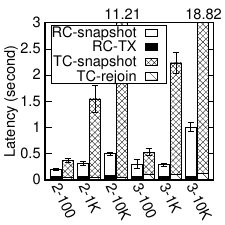}
		\vspace{-0.6em}
		\caption{Latency}
		\label{fig:mergelatency}
	\end{subfigure}
	\caption{Merge performance. (a) Throughput of two (left) and three (right) 3-node subclusters merging 
into one cluster. (b) Merge latency for \sysname{} (RC) and TiKV/CockroachDB emulation (TC). 
	The x-tic, $a$-$b$, indicates $a$ subclusters with $b$ KV pairs merging.}
	\vspace{-1em}
\end{figure}

\subsection{Merge Performance}
\label{subsec:evalmerge}

\para{Througput.}
We merge two and three 3-node clusters into a single 6-node (left) and 9-node
(right) clusters, respectively, and measure the throughput
(Figure~\ref{fig:mergetime}). We use a small amount of load (i.e., 2 clients 
with uniform random puts) as it is desired to merge clusters when the nodes
are underutilized. The merge is blocking so there is a small throughput dip 
when merging happens at the 30-second mark. Individual clusters before and after 
the merge perform almost the same due to etcd's extended wait time for batching 
under a low load: the average latency for each request doubles after the merge.


\para{Latency.}
For a latency evaluation, we compare \sysname{} against TC that coalesces all 
subcluster data in one of the subclusters, terminates all subclusters but the 
one with the coalesced data, and adds all nodes from terminated subclusters to
the live one. 

We use the same node configurations as the throughput experiment but with etcd
instances holding 100, 1K, and 10K KV pairs. The 2PC takes almost constant time
for \sysname{} across different settings (Figure~\ref{fig:mergelatency}). 
For both \sysname{} and TC, 
merging the data takes the longest. Especially, TC exchanges data with 
more blocking and can take from 1.7x to 20x more time than \sysname{} depending 
on the data size. Similar to the split, the \sysname{} approach is more robust and 
performant. 

\revfaileval{
\subsection{Fault Tolerance During Split and Merge}
\label{subsec:evalfault}
We analyze the fault-tolerance of \sysname{} and TC deployments in Sections 
\ref{subsec:evalsplit} and \ref{subsec:evalmerge} based on the minimum number 
of node failures that can completely stop the split and merge
(\cref{tbl:failure}).
For TC, we consider a non-replicated cluster manager (CM) and a 
replicated CM (CM-repl) on Raft. The fault tolerance of \sysname{} 
varies by the operation phases, but only cluster granularity
interrupts can stop the operation.}

\revfaileval{
For split, $C_{old}$ needs to fail to stop \splitenterjoint{} (phase 1), and 
all $N$ subclusters have to fail to stop \splitleavejoint{} (phase 2) 
as subclusters can leave joint mode on their own. Merge needs every subcluster
during the 2PC (phases 1 and 2) and the data exchange (phase 3), so 
disabling one subcluster in any phase is enough to stop the operation. 
For TC, one can fail the subclusters as in \sysname{}, but 
failing the non-replicated CM or the CM cluster ($C_{cm}$) is enough. TC with
non-replicated CM is the least fault tolerant among all when $C_{sub}$ 
consists of at least three nodes (i.e., $f_{sub} \ge 1$). If the sizes of 
$C_{sub}$ and $C_{cm}$ are the same (e.g., 3), \sysname{} can tolerate more 
failures for split and the same number for merge compared
to TC. However, recall that \sysname{} does not need separate CMs, is more
concurrent, and blocks minimally.}


\begin{table}[t]
    \centering
    \footnotesize
    \begin{tabular}{c|ccc|cc}
        \hline
        N-way& \sysname{}& \sysname{} & \sysname{} & TC & TC \\
	    operation& Phase 1 & Phase 2 & Phase 3 & CM & CM-repl. \\
        \hline
        Split & $f_{old}+1$ & $N(f_{sub}+1)$ & - & $1$ & $f_{cm}+1$ \\
        Merge & $f_{sub}+1$ & $f_{sub}+1$ & $f_{sub}+1$ & $1$ & $f_{cm}+1$ \\
        \hline
    \end{tabular}
    \caption{\revfaileval{Minimum number of node failures to completely stop the operation
    under a uniform subcluster sizes.
	$f$ is the number of node failures a \sysname{}/Raft/etcd cluster can tolerate ($old$: 
    initial configuration; $sub$: subcluster configuration; $cm$: replicated 
	cluster manager configuration).}}
	\vspace{-1em}
    \label{tbl:failure}
\end{table}

\subsection{Membership Change}
The performance of the AR-RPC, the JC, and \sysname{} membership change 
is dominated by the number of consensus steps analyzed in 
Section~\ref{subsec:membercomparison}. The consensus step to commit a command 
(e.g., a new configuration) for AR-RPC, JC, and \sysname{} takes on average 
11.4 ms. \sysname{} performs equal to or better than AR-RPC and JC 
for membership changes under practical cluster sizes between 2 to 5 except for 
when reducing the cluster size from 5 to 2, \revb{which requires one extra 
consensus step than JC.}

%
%
%
%
%
%
%
%

\section{Related Work}

\para{Sharding SMR.}
TiKV~\cite{tidb} and CockroachDB~\cite{cockroachdb} use the multi-Raft 
design with logical cluster split and merge features (\cref{subsec:tikvsm}). 
Different from \sysname{}, their cluster split and merge operations are driven 
by the cluster manager, which uses sophisticated distributed locks and 
transactions and could become a single point of failure. One may choose to 
replicate the cluster manager for fault tolerance (e.g., as in
Section~\ref{subsec:evalfault}), but it adds a burden to 
manage another cluster. \sysname{} relies on the existing consensus mechanism of 
participating clusters which is self-contained, more robust, and simple. 
Note that \sysname{} protocol can be complementarily added to the multi-Raft 
design for the logical Raft instance management.



Logically partitioning the SMR logs with some overlay logic on a single SMR 
cluster is another approach for scaling the SMR approach~\cite{tango, fuzzylog,
delos, matchmaker-paxos}. \sysname{} works at the underlay SMR log and is a
complementary design.
 
Blockchains are SMR logs in Byzantine environments and a few block chain designs
employ sharding~\cite{gearbox, omniledger, rapidchain, brokerchain}. However, 
their designs and assumptions surrounding reconfigurations are different from 
those of \sysname{}, as blockchains heavily rely on probability and assume  
fluctuating memberships.

\para{Membership Change Schemes for Consensus-based 
Distributed Systems.}
\revb{Consensus and similar linearizable distributed
systems~\cite{vivaladifference} have been at the center of managing 
other system configurations and their own reconfiguration has been 
focused on the single cluster membership change. 
Most of these reconfiguration 
protocols~\cite{paxos-moderate, stoppablepaxos, reconfiguringpaxos, 
generalizedconsensus, wormspace, smart, vrrevist, zkreconfig, bftsmart} entail 
blocking until the configuration fully commits or require administrator
interventions.} Except for the merge operation, \sysname{} applies the 
reconfiguration in a wait-free fashion so other requests can be serviced 
during the reconfiguration steps.

\para{Membership Change Schemes Based on Atomic Registers.}
\camready{DynaStore~\cite{dynastore} and RAMBO~\cite{rambo,rambo2} propose cluster 
reconfiguration protocols using distributed atomic registers (e.g., ABD protocol~\cite{swmr}). 
Atomic registers commit writes atomically using quorums, but they are free from 
FLP impossibility~\cite{flp}: while the register is strongly consistent and 
highly available, the ordering of concurrent writes to it is difficult to control. 
Thus, membership services use multiple registers that apply partially ordered deltas 
to existing configurations, and merging the information in the registers yields
the final configuration. However, due to the quorum-based nature, these services and 
\sysname{} share commonalities which include the configuration overlap between
consecutive configurations (e.g., as in quorum overlap in \sysname{}) and the need for
consulting multiple configurations (e.g., as in joint quorums in \sysname{}) during the 
reconfiguration process.}

%
%


\para{Pulling Data.} \sysname{}'s log pulling during the split and merge is 
inspired by the leader election phase of Paxos~\cite{paxos-parliament, paxos-simple}. 
While \sysname{} only pulls committed entries for recovering nodes, a closer 
idea to Paxos of pulling uncommitted log entries is applied to a Raft design in 
MongoDB~\cite{mongodb}. \revb{Mirador uses pulling for enhanced modularization 
and debugging: it leverages pull-based micro-replication to
build a Byzantine fault-tolerant SMR protocol~\cite{microrep}.} 

\para{Formal Proofs.}
Due to their sophisticated nature, Raft and Paxos 
protocols are formally verified in many studies~\cite{ado, wormspace, 
ironfleet, cppraft, paxosepr, sdpaxos}. \camready{The \sysname{} proof relies 
on the overall safety proof of Raft~\cite{cppraft} and Adore's~\cite{adore} 
proof of Raft membership changes, which are mechanized using Rocq.}

\section{Conclusion}

We present \sysname{}, a reconfigurable Raft, with a new reconfiguration 
protocol that splits and merges Raft clusters and efficiently changes the 
cluster membership. The split and merge protocol is the first self-contained 
and fault-tolerant design based on the consensus of all relevant subclusters. 
The \sysname{} membership change scheme is more efficient than Raft's approach 
for practical cluster sizes. We present proof of \sysname{}'s safety and 
liveness and demonstrate the efficiency of an etcd-based implementation in a 
public cloud.

\section*{Acknowledgments}
We thank the anonymous reviewers for their insightful reviews and 
Jos\'{e} Orlando Pereira for shepherding the paper. We also thank Pruthvi Prakash Navada for testing and packaging the artifact. The artifact (i.e., \sysname{} code and Rocq proof sketches) is available at \url{https://zenodo.org/records/15283088}. This work is supported by the Khoury Apprenticeship Program, NSF award \#2019285, and the National Research Foundation of Korea (NRF) grant funded by the Korean government (MSIT) (RS-2025-00556905).

%

\balance
\bibliographystyle{IEEEtran}
\bibliography{refs, adore-refs}

\pagebreak

\appendixver{\appendix
\section{Safety and Liveness Proofs}\label{app:proofs}

\renewcommand{\disj}{\mathbin{\#}}
\renewcommand{\ndisj}{\mathbin{\centernot\#}}
\renewcommand{\lastTerm}{\operatorname{lastTerm}}
\renewcommand{\length}{\operatorname{length}}

\subsection*{Safety proofs}

With the approach described in Section~\ref{sec:proof}, three of the listed lemmas - LO (Leader Append-Only), ES (Election Safety), and LM (Log Matching) - are either identical or nearly identical to the corresponding proofs in the original Raft.

\begin{proof}[Proofs for LA, ES, and LM]
With the epoch number fixed, the proofs closely mirror those of the original Raft algorithm. Each relies on basic quorum-overlap arguments as the core decision procedure. Because both log entry confirmation and leader election decisions are made through quorum agreement, the corresponding safety properties are preserved. The following sections present detailed proofs for each of these properties.
\begin{enumerate}
\item \textbf{Leader Append-Only}:  
Once a server becomes leader in a given epoch, it only appends new entries to its log and never overwrites or deletes existing entries. This property holds in \sysname{} because the leader exclusively controls log replication for its epoch, and the log update mechanism enforces that all entries are appended to the end of the log. This directly mirrors the append-only guarantee of Raft, where leaders never modify previously committed log entries.

\item \textbf{Election Safety}:  
With the epoch number fixed, elections in \sysname{} proceed either through a normal quorum or a joint quorum. Since every joint quorum includes at least one normal quorum as a subset, and since each server votes at most once per epoch, two different leaders cannot be elected in the same epoch. This preserves the core Election Safety invariant of Raft: at most one leader can be elected per term. The argument follows the same structure as in Raft—vote uniqueness, quorum intersection, and monotonic term progression ensure safety even with the generalized quorum mechanism.

\item \textbf{Log Matching}:  
With a fixed epoch, \sysname{} preserves the Log Matching Property: if two logs contain an entry with the same index and term, then the logs are identical in all preceding entries. The proof closely follows the original Raft argument. Log replication in \sysname{} ensures that followers only accept entries that match their existing logs at the given index and term. This prevents divergence and guarantees that logs with a common suffix are consistent in their prefixes as well. As in Raft, this property plays a key role in ensuring state machine consistency.
\end{enumerate}
\end{proof}	

However, leader completeness (Definition~\ref{def:leader-completeness}) requires significant alteration compared to the original Raft due to the modifications introduced by \sysname{}, and the proof is described in Section~\ref{sec:proof}. Within the proof, we use Definition~\ref{def:cluster-well-formedness} as part of it.
The main purpose of Definition~\ref{def:cluster-well-formedness} is showing that all the cooperating nodes of that epoch share the cluster configuration.
For example, consider a situation where a follower $a$ of epoch $e$ and cluster configuration $C_a$
accepted a message from a leader $p$ with cluster configuration $C_p$.s
The following entails detailed proofs related to Definition~\ref{def:cluster-well-formedness} (Cluster Well-Formedness).
%
%

\begin{proof}
The proof is conducted through strong induction on the epoch number.
Suppose there are nodes $a_1$ and $a_2$ with the same epoch number $e$, where their assigned clusters are $C_1$ and $C_2$, respectively.
In this case, $C_1 = C_2 \vee C_1 \disj C_2$ when $\disj$ stands for ``disjoint'' set.
  \begin{itemize}
    \item If $C_1 \disj C_2$, then the goal immediately holds.
    \item
      If $C_1 \ndisj C_2$, let's select a node $a$ that satisfies $a \in C_1 \cap C_2$.
      Find a chain of reconfigurations $C_{10} \to C_{11} \to \cdots \to C_{1m} = C_1$ and $C_{20} \to C_{21} \to \cdots \to C_{2n} = C_2$ where
      $C_{10}$ and $C_{20}$ are initial clusters, and $a$ is a member of all the constituting clusters of the two chains.
      With the fact of the initial condition, 
      $C_{10} = C_{20} \vee C_{10} \disj C_{20}$ holds, 
      but the second case is a contradiction since $a \in C_{10}$ and $a \in C_{20}$.
      Thus $C_{10} = C_{20}$.
      By using  Definition~\ref{def:log-consistency} with induction hypothesis (Cluster Well-Formedness) on epoch 0,
        we know that Definition~\ref{def:log-consistency} holds on epoch 0.
      From this, there is only one possible reconfiguration for $C_{10} = C_{20}$.
      Since $C_{10} \to C_{11}$ and $C_{20} \to C_{21}$, $C_{11}$ and $C_{21}$ are resulting clusters of the same reconfiguration.
      But then $C_{11} = C_{21}$, since resulting clusters of a reconfiguration must be disjoint each other, and $C_{11} \ndisj C_{21}$ because of $a$.
      Also, the induction hypothesis on epoch 1 can derive the fact, $C_{12} = C_{22}$, via a similar process. 
      Repeating this process, we can show that $m = n$ and $C_{1m} = C_{2n}$. Thus $C_1 = C_2$.
  \end{itemize}
Therefore, Definition~\ref{def:cluster-well-formedness} holds.
\end{proof}

With all ingredients discussed, we finally prove Definition~\ref{def:log-consistency} (Log Consistency).
%
\begin{proof}
  Let $i \mapsto (x_1, t_1)$ and $i \mapsto (x_2, t_2)$ be the two committed entries.
  Then there must be directly committed entries $i_1' \mapsto (x_1', t_1')$ and $i_2' \mapsto (x_2', t_2')$ that lead to the commit of each $i \mapsto (x_1, t_1)$ and $i \mapsto (x_2, t_2)$.
Utilize Definition~\ref{def:log-matching} and Definition~\ref{def:leader-completeness} with directly committed entries, and we can conclude that $(x_1, t_1) = (x_2, t_2)$.
\end{proof}

Now we show that all sub-definitions that are necessary to show our main safety statement,
but most of them still rely on Definition~\ref{def:cluster-well-formedness}. 
Therefore, we show Definition~\ref{def:cluster-well-formedness}.
Indeed, the proof structure is mutually recursive;
Definition~\ref{def:log-consistency} of epoch $e$ depends on Definition~\ref{def:cluster-well-formedness} of epoch $e$,
and Definition~\ref{def:cluster-well-formedness} of epoch $e$ depends on Definition~\ref{def:log-consistency} of all the previous epochs. 

By combining Definitions~\ref{def:cluster-well-formedness} and~\ref{def:log-consistency}, we can prove that Definition~\ref{def:log-consistency} holds on every epoch.
The top level theorem, state machine safety (Theorem~\ref{thm:state-machine-safety}), is a corollary of this fact.

\subsection{Proof mechanization}
As mentioned in Section~\ref{subsec:proof-mech}, we have structured the proof using Rocq to validate our high-level proof structure and mitigate the risk of human errors in this and other complex parts of the system. We have intentionally omitted several trivial or tedious subproofs, such as the monotonicity of increasing term and epoch numbers for a single node.

Using Coq for this purpose presents both advantages and challenges. On one hand, mechanization provides a high degree of confidence in the correctness of our reasoning. On the other hand, it introduces additional complexity to the proof process, as it requires the development of multiple formal definitions that precisely capture the system state and are expressed using Coq definitions with several auxiliary annotations.

	Our long-term goal is to achieve a higher level of assurance than the current version. In particular, we aim to complete all the tedious and detailed invariant proofs that are currently marked with \texttt{admit} in our Coq development. Completing these will yield a fully verified proof with no remaining gaps.
}{}

\end{document}